\documentstyle[aps,twocolumn,epsfig]{revtex}
\newcommand{\bg}[1]{\mbox{\boldmath$#1$}}
\newcommand{\mdue}{|2\rangle}
\newcommand{\muno}{|1\rangle}

\begin{document}

\draft

\title{Time-domain atom interferometry across the threshold for Bose-Einstein
  condensation} \author{F.~Minardi, C.~Fort, P.~Maddaloni\cite{pm},
  M.~Modugno, and M.~Inguscio}
  \address{INFM - European Laboratory for
  Non linear Spectroscopy (LENS) and Dipartimento di Fisica,\\
  Universit\`a di Firenze, Largo E. Fermi 2, I-50125 Firenze, Italy.}

\date{\today}

\maketitle

\begin{abstract}
  
  We have performed time-domain interferometry experiments with matter
  waves trapped in an harmonic potential above and below the
  Bose-Einstein phase transition.  We interrogate the atoms according
  to the method of separated oscillating fields, with a sequence of
  two radio-frequency pulses, separated by a time delay $T$. We
  observe the oscillation of the population between two internal
  Zeeman states, as a function of the delay $T$.
  
  We find a strong depletion of the interference fringes for both the Bose
  condensates and the thermal clouds above condensation, even at very short
  times, when the clouds are still overlapping. Actually, we explain
  the observed loss of contrast in terms of phase patterns imprinted
  by the relative motion, as a consequence of the entanglement
  between the internal and external states of the trapped atoms.

\end{abstract}

The question about the coherence of Bose-Einstein condensates (BECs)
\cite{1stBEC}, and the characterization of their phase properties have drawn
considerable attention in the recent literature. 

The first evidence of a definite phase for weakly interacting
condensates dates back to the experiment of the MIT group \cite{intketterle},
where high contrast matter-wave interference fringes were observed in the 
density distribution of two freely expanding BEC's. 
Subsequent experiments performed at JILA \cite{phaseJILA,phaseJILAcondmat} 
measured the relative phase of two condensate in different internal 
(hyperfine) states experiencing almost the same trapping potential.  

Other experiments have further investigated this subject
\cite{PhillipsPRL83,JapanMZ,phasePhillips,Ertmerwaveguide}.    
In particular, the group of W.~Phillips has recently measured the evolution of
the spatial profile of the phase of BEC's, by using a Bragg-interferometer
\cite{phasePhillips}. They have shown that a trapped condensate is
characterized by a uniform phase, and after release from the trap 
develops a non-uniform phase profile. 

In a recent Letter \cite{PRL} we have demonstrated an experimental
method for a sensitive and precise investigation of the interaction
between two Bose-Einstein condensates \cite{PRA}. In this paper we
present an interferometry experiment performed in the time-domain on
the same system, which allows to gain a deeper insight into
Ramsey interferometry \cite{Ramsey}
 with ultra-cold atoms across the BEC phase transition.

It is well known that Ramsey fringes are readily observable with a
sample at room temperature. The reason is that Ramsey signals rely
on the persistence of coherence between two distinct atomic levels,
no matter what the state of the particles center-of-mass is. This is true
only when the internal and external degrees of freedom are
uncoupled. Whenever the latter condition fails, a depletion of the
Ramsey fringes visibility can occur and has indeed been observed
\cite{separatedRamsey}.  Thus, the entanglement of external and internal
states of atoms trapped in a magnetic potential is the basis for the
use of Ramsey method to characterize the phase properties of a
condensate and a thermal cloud near BEC. In the JILA experiment \cite{phaseJILA}, such
an entanglement caused the observed Ramsey fringes in the population
of the two hyperfine levels to undergo a loss of contrast, as a
consequence of the reduced spatial overlap between the two BEC's, due
to their mutual repulsion. 

In our experiment we assist to a similar
depletion but on a much faster time-scale, of the order of tens of
$\mu$s, that cannot be explained with the reduction of spatial
overlap.
We prepare, and subsequently probe the system
with a sequence of two identical radio-frequency (rf) pulses, separated by a delay
time $T$.  The relative phase accumulated by the condensates produces Ramsey fringes in the population of each
level, as a function of the delay $T$.  We observe a strong reduction of
the oscillation amplitude even at very short times, when the condensates are
still almost completely overlapping, due to the relative velocity
acquired by the condensates.
We compare this reduction with that of a cloud of thermal atoms, at a
temperature three times larger than the phase-transition temperature $\Theta \sim 3 \Theta_c$.

We prepare a condensate of typically $2 \times 10^5~^{87}$Rb atoms in
the $|F=2,m_f=2\rangle$ hyperfine level, confined in a 4-coils
Ioffe-Pritchard trap elongated along the $z$ symmetry axis \cite{epj}.
The axial and radial frequencies for the $\mdue$ state are $\omega_{z
  2}= 2 \pi \times 13$~Hz and $\omega_{\perp 2}= 2 \pi \times 131$~Hz,
respectively, with a magnetic field minimum of $2.86$~Gauss. Then, we
apply an rf-pulse to split the initial condensate into a
coherent superposition of different Zeeman $|m_f\rangle$ sublevels of
the $F=2$ state.  The atoms transferred in a different sublevel move
away from the $\mdue$ equilibrium position with an acceleration that
depends on their $m_f$ value.  Thus, the state of motion becomes
entangled with the internal atomic state.

In this experiment we use a 24 cycles rf-pulse at 2~MHz, which quickly leaves the $|2\rangle$ state with 
$\sim 41$\% of the initial number of atoms, transferring an equal 
part of atoms ($\sim 41$\%) to the $|1\rangle$ state and 
$\sim 15$\% of the atoms the $|0\rangle$ state, 
$\sim 3$\% to the $|-1 \rangle$ and $\sim 0$\% to the $|-2 \rangle$ states.

Even though all the five levels should be taken into account in the
early stages of the evolution (before the atoms in the $|0 \rangle$,
$|-1 \rangle$ and $|-2 \rangle$ states leave the trap), the basic
features can be explained by considering only the dynamics of the two
most populated levels, {\em i.e.} $|1 \rangle$ and $|2 \rangle$.  We
will discuss later how the other three levels affect the overall
behaviour of the system.

We describe our double condensate system by a spinor wavefunction, where the
upper and lower components refers to states $\mdue$ and $\muno$,
respectively.
Immediately after the first rf-pulse, the wavefunction can be written
as:

\begin{equation} 
\Psi(\bg{r};t=0) = \frac{\sqrt{N_0} u(\bg{r};0)}{\sqrt{2}} \left(
\begin{array}{c}
1\\
i 
\end{array}
  \right) 
\label{eq:t0}
\end{equation} 
where $\sqrt{N_0} u(\bg{r})$ is the equilibrium wavefunction of the
$N_0$ atoms of the initial $\mdue$ condensate, in the Thomas-Fermi
(TF) regime \cite{bec_review}.  As the rf-pulse is much shorter than
the oscillation periods of the harmonic trap, we could safely take the
spatial wavefunction $u(\bg{r})$ to remain unchanged and flop only the
internal state.

In the subsequent
free evolution, the relative phase between the two components
accumulates with a rate proportional to the difference of chemical
potentials $\mu_2 - \mu_1$. Moreover, $u(\bg{r};0)$ is no longer the
equilibrium wavefunction for $\mdue$ nor for $\muno$: the spatial
wavefunctions evolve as dictated by two coupled GP equations
\cite{PRA} into $u(\bg{r};t)$ and $v(\bg{r};t)$, respectively.

By applying a second rf-pulse, identical to  the first, after a
time delay $T$, we suddenly mix the two components
\begin{equation} 
\psi(\bg{r};T) = \frac{\sqrt{N_0}}{2} 
\left(
\begin{array}{c}
u(\bg{r};T) e^{-i \omega_0 T} + i\, v(\bg{r};T) \\
i\, u(\bg{r};T)\,e^{-i \omega_0 T}  + v(\bg{r};T) 
\end{array}
  \right)
\label{eq:tT2}
\end{equation} 
with $\omega_0=(\mu_2-\mu_1)/\hbar$.

We then separate the two internal states and
independently count the number of atoms in each, given by the
integrated square
modulus of the corresponding components, showing an oscillating
behaviour at the frequency $\omega_0$ \cite{whyomega}.
 
For times $T$ much shorter than the harmonic periods, we can neglect all
but the most relevant effect of the spatial wavefunctions evolution:
due the differential gravitational ``sagging'', the $\muno$ condensate
acquires a downward time-dependent momentum $-\hbar q(t)$.
In particular, we will also neglect the loss of spatial overlap
arising from the relative displacement.
Then, taking $u(\bg{r};T)=u(\bg{r};0)$, $v(\bg{r};T)=i u(\bg{r};0)\exp(-iq(T)y)$ and given
the normalization $\int |u(\bg{r};0)|^2 dy =1$, we have
\begin{eqnarray} 
N_2 (T)&=& 
N_0 \frac{1}{2} [1 - A_c(T) \cos(\omega_0 T)] \nonumber \\
N_1(T)&=&N_0-N_2(T)
\end{eqnarray}
with the slow time-dependent oscillation amplitude 
\begin{eqnarray} 
A_c(T)&=& \int |u(\bg{r};0)|^2 \cos(q(T) y) d\bg{r}.
\label{eq:cond-ampl}
\end{eqnarray} 
The amplitude of the quadrature component vanishes because
$|u(\bg{r};0)|^2$ is an even function of the $y$ coordinate.

The amplitude $A_c(T)$ decays as the relative velocity increases and
$q^{-1}$ becomes of the order of the vertical extension of the
original condensate. Eventually, Ramsey fringes are completely washed
out. The relative displacement would give the same result, but only at
later times.

\begin{figure}
\epsfig{file=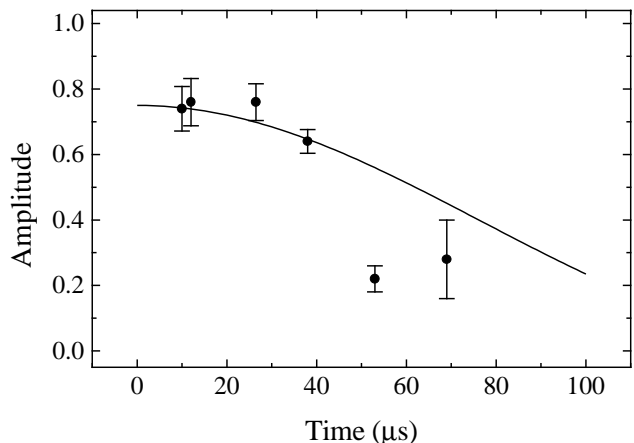,width=8.5cm}
\caption{Peak-to-peak amplitude of the oscillating observable
  $N_2/(N_2+N_1)$, i.e. the $\mdue$ condensate
  population normalized to the total number of atoms in $\mdue$ and
  $\muno$: experiment and calculation (solid line)} \label{fig:con-ampl}
\end{figure}

In Fig.~\ref{fig:con-ampl} we plot the experimental data of the
oscillation amplitudes for the fraction $N_2/(N_1+N_2)$ versus the
time delay $T$.  To obtain each data point, we have sampled one
oscillation period around the reported $T$ value. We have taken about
10 points per period and each point is the average of few
acquisitions.  
To
compare the experimental points with the values predicted by
Eq.~(\ref{eq:cond-ampl}), we rescale the latter by a suitable factor, chosen
to match the data around $T=0$. As for the damping time, we find a
satisfactory agreement between theory and experiment.

Actually, for the double condensate we can refine the above model to
include the mean-field repulsion in the wavefunctions evolution: to
this end, we numerically integrate two coupled Gross-Pitaevskii (GP)
equations, according to the model described in Ref. \cite{PRA}.  On
general ground, in the first stages of the evolution the phase of each
component can be written as a quadratic form in the spatial
coordinates \cite{phasePhillips,bec_review}, describing the mean-field
expansion/contraction of the condensates in the Thomas-Fermi regime
\cite{bec_review} and the motion along the vertical direction $y$. The
results of the numerical simulations, in the time range considered
here, show the following: {\em (i)} the phase of the $|2\rangle$
condensate remains almost uniform, and only for later times ($t\simeq
0.5$~ms) develops a negative curvature, due to the fact that the
condensate is contracting after the initial transfer of atoms to the
other levels; {\em (ii)} the phase of the $|1\rangle$ condensate is
dominated by a linear term
\begin{equation}
\phi_1(\bg{r},t) \simeq  {m\over\hbar} (v_0(t) + \delta v(t))y
\label{eq:phase1}
\end{equation}
where $v_0(t)$ is the velocity acquired during the fall in the trapping
potential
\begin{equation}
v_0(t) = -{g\over \sqrt{2}\,\omega_{\perp 2}}\sin\left({\omega_{\perp 2}t\over
\sqrt{2}}\right) 
\label{eq:vel}
\end{equation}
and $\delta v(t)$, which is negligible for $t<0.5$~ms, is due to the
mutual repulsion between the two condensates.  Thus, the GP
simulations confirm that, at short times, the simple model adopted
above correctly describes the basic features of the system.  

\begin{figure}
\epsfig{file=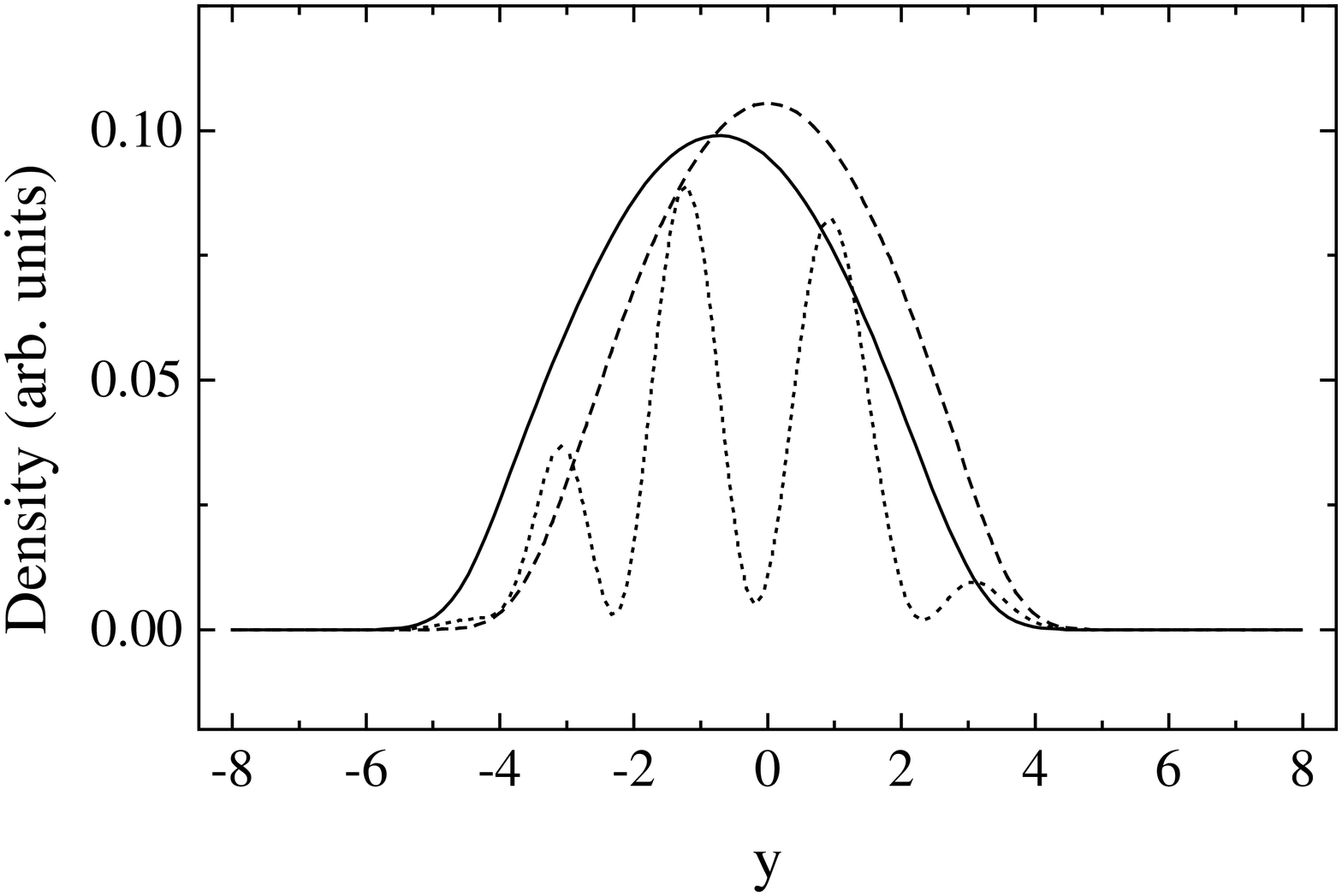,width=8.5cm}
\caption{Overlap of the two condensates at the second
  rf-pulse. Density profiles along the vertical axis $y$ of condensates $\mdue$
  (dashed-line) and
  $\muno$ (solid line) before
  the pulse, and of condensate $\muno$ (dotted line) immediately after.
The curves are obtained by solving the GP equations for
  the two-state model in Ref.  \protect\cite{PRA}.  Lengths are given
  in units of $a_{\perp2}=
  [\hbar/(m\omega_{\perp2})]^{1/2}=0.94~\mu$m; $T=0.5~ms$.}
\label{fig:overlap}
\end{figure}

The phase term (\ref{eq:phase1}) is responsible for 
washing out completely the Ramsey fringes even at very short times,
when the condensate are still almost completely overlapping (as shown
in Fig.~\ref{fig:overlap} the overlap between the two wavepackets is
substantial even at $T$ as long as 0.5~ms). Physically, across the
spatial extension of the wavepacket there are regions of alternated
positive and negative interference, with a vanishing net transfer of
atoms.

For a thermal cloud we can consider the system to be in a given
quantum state and then take an ensemble average over all the
accessible states. This way, we need only to replace the
time-dependent amplitude (\ref{eq:cond-ampl}) with
\begin{equation} 
A_{\rm th}(T)= \frac{1}{N_0} \sum_{ \{n\}}  \int f_{\{n\}} 
\cdot |\Psi_{\{n\}}(\bg{r})|^2 \cos(q(T) y) d\bg{r}  
\label{eq:th-ampl}
\end{equation} 
where $f_{\{n\}}=[\exp((\epsilon_{\{n\}} - \mu)/k\Theta) -1]^{-1}$ is
the Bose mean occupation number of the harmonic trap eigenstate
$\Psi_{\{n\}}=\psi_{n_1}(x)\psi_{n_2}(y)\psi_{n_3}(z)$ with energy
$\epsilon_{\{n\}}$, $\mu$ is the chemical potential, given by the
normalization $\sum_{\{n\}} f_{\{n\}}=N_0$, and $\Theta$ the
temperature.  By carrying out the integration over $x,z$ and the sum
over the corresponding quantum numbers $n_1,n_3$, we find
\begin{eqnarray} 
A_{\rm th}(T)&=& \frac{(k\Theta)^2}{N_0\hbar^2 \omega_x \omega_z }
\sum_{n_2} g_{2} ( \exp(\frac{\mu-n_2\hbar\omega_y}{k\Theta} ) ) \times
\nonumber\\ &&\times\int |\psi_{n_2}(y)|^2 \cos(q(T) y) dy,
\label{eq:th-ampl2}
\end{eqnarray} 
the $g_2(x)=\sum_{l=1}^{\infty} x^l/l^2$ function being the result of
replacing the discrete sum over $n_1,n_3$ with a double integral.
In principle, one should allow for the rf-pulses to act differently on
the different harmonic oscillator eigenstates as the detuning
varies. This would introduce $\{n\}$-dependent weights in
Eq.~(\ref{eq:th-ampl}). However, we have verified that for the
relevant levels at $\Theta\simeq 0.4~\mu$K, these weights are equal
within a few percent.

We note here that Eq.~(\ref{eq:th-ampl}) can be rewritten
\begin{equation} 
A_{\rm th}(T)= \frac{1}{N_0} \int n(\bg{r}) \cos(q(T) y) d\bg{r} 
\label{eq:th-ampl3}
\end{equation} 
where $n(\bg{r})$ is the spatial density.
In this form, it is evident that the thermal cloud behaves as a
coherent wavepacket, the reason being that the displacement occurring
between the two rf-pulses is much less than the thermal coherence
length ($\lambda_{\rm th}=h/\sqrt{2\pi mk\Theta}$).

\begin{figure}
\epsfig{file=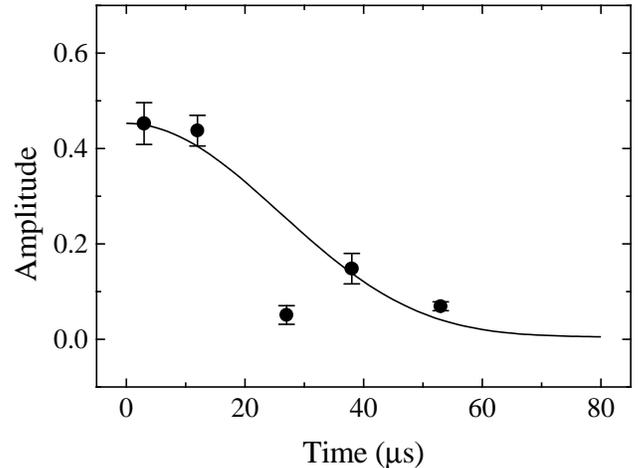,width=8.5cm}
\caption{Peak-to-peak amplitude of the oscillating observable
  $N_2/(N_2+N_1)$, for thermal atoms at $\Theta=0.4~\mu \rm{K} \simeq
  3\Theta_c$: experiment and calculation (solid line)}
\label{fig:th-ampl}
\end{figure}

We point out that for the non-condensed sample, where typical densities are a
factor 30 lower than those of BEC, we neglect the atom-atom
interactions, which justifies the above single-particle analysis.  In Fig.~\ref{fig:th-ampl} we plot the amplitude of the
observed $N_2/(N_2+N_1)$ oscillation for several values of the time
delay $T$ between the two rf-pulses and we compare with the
corresponding predictions given by Eq.~(\ref{eq:th-ampl}), rescaled to
match the experimental values for $T\rightarrow 0$. As for the damping
time, we have again a satisfactory agreement; thus, we believe that the
cause of the Ramsey fringes loss of contrast is well understood.

However, the experimental values are about a factor 2 less than those
given by Eq.~(\ref{eq:th-ampl}).  
As we already noticed, part of the atoms end up in the $m_f=0,-1,-2$ Zeeman
sublevels after the second rf-pulse. Two are the main consequences:
$(i)$ the peak-to-peak oscillation amplitude of the $N_2/(N_1+N_2)$
fraction cannot exceed 0.87, both for the condensates and the thermal
clouds; $(ii)$ the oscillation waveforms deviate from a pure
sinusoidal behaviour.

To conclude, we point out that our system is suitable to study the
possibility of a revival of the Ramsey fringes after the condensates
have been spatially well separated. As the center-of-mass of the
$\muno$ condensate undergoes harmonic oscillations around its
equilibrium position \cite{PRL,PRA}, it comes back to rest at its
initial position. According to the above description one should expect
a revival of the oscillations in the relative population when the
condensate $|1\rangle$ and $|2\rangle$ come to overlap again with
almost vanishing velocity.  The numerical solution of the GP equations
of the two-level model shows indeed that over time scales of tens of
ms the Ramsey fringes are characterized by collapse and revival \cite{cargese}.

To summarize, we have studied the Ramsey interference of atomic
clouds across the BEC phase-transition, with a system where the two
involved states have equilibrium positions far apart.  We have
shown that the phase pattern imprinted on the moving wavepacket by its
acquired velocity washes out the Ramsey oscillations of the fractional
populations well before that the spatial overlap decreases.  By
repeating the same experiment on a thermal cloud at three times the
condensation temperature, we have observed that the same mechanism is
responsible for an even faster damping of oscillation contrast.  In
this respect, there is no substantial difference from a thermal cloud
and a Bose condensate. The analogy with light optics is
straightforward: indeed, to observe interference between two paths we
need only the difference between the paths length not to exceed the
coherence length.

We thank T.~Lopez-Arias, A.~Trombettoni, A.~Smerzi for helpful
discussions and M.~Artoni for a careful reading of the manuscript.
This work was supported by Cofinanziamento MURST and EU under contract
No. HPRICT1999-00111.



\begin{references}
\bibitem{whyomega} 
We note that, unlike in usual Ramsey experiments where the signal
oscillates at the frequency difference between the atomic
frequency and the interrogation frequency, here we record oscillations
at the atomic frequency. The reason is that the phase of the
second pulse remains the same, no matter what is the time delay $T$.
Thus, varying $T$ we just scan the phase of the atomic dipole.
\bibitem{cargese} We defer a more complete discussion on this point
  to a forthcoming publication.
\end{references}
\end{document}